\documentclass[twocolumn,prd,amsmath,amssymb]{revtex4}

\usepackage{ifthen}
\usepackage{graphicx}
\usepackage{dcolumn}
\usepackage{bm}
\usepackage{xspace}
\usepackage{mathrsfs}
\usepackage{amsmath}
\usepackage{amssymb}
\usepackage{amsfonts}
\usepackage{dcolumn}

\newcommand{\ee}{\ensuremath{e^{+}e^{-}}\xspace}
\newcommand{\mumu}{\ensuremath{\mu^{+}\mu^{-}}}
\newcommand{\JP}{\ensuremath{J/\psi\,}\xspace}
\newcommand{\PP}{\ensuremath{\psi'}\xspace}
\newcommand{\VEPP}[1]{\mbox{VEPP-#1}\xspace}
\newcommand{\Lum}{\ensuremath{L}\xspace}
\newcommand{\Ilum}{\ell\xspace}

\begin{document}

\title{ New precision measurement of the \JP- and \PP-meson
  masses\footnote{Partially supported by the Russian Foundation for
    Basic Research, Grants 01-02-17477, 02-02-16321, 02-02-17321 and
    the Presidential Grants 1335.2003.2, 1346.2003.2 for support of
    Leading Scientific Schools.  } 
}

\author{V.M.~Aulchenko}
\author{S.A.~Balashov} 
\author{E.M.~Baldin} 
\thanks{e-mail: E.M.Baldin@inp.nsk.su (corresponding author)}
\author{M.Yu.~Barnyakov} 
\author{S.E.~Baru} 
\author{I.V.~Bedny}
\author{O.L.~Beloborodova}
\author{A.E.~Blinov}
\author{V.E.~Blinov}
\author{A.V.~Bogomyagkov} 
\author{A.E.~Bondar} 
\author{D.V.~Bondarev}
\author{A.R.~Buzykaev}
\author{S.I.~Eidelman}
\author{V.R.~Groshev}
\author{S.E.~Karnaev}
\author{V.A.~Kiselev}
\author{S.A.~Kononov}
\author{K.A.~Kotov}
\author{E.A.~Kravchenko}
\author{E.V.~Kremyanskaya}
\author{E.B.~Levichev}
\author{V.M.~Malyshev}
\author{A.L.~Maslennikov}
\author{O.I.~Meshkov}
\author{S.E.~Mishnev}
\author{N.Yu.~Muchnoi}
\author{A.I.~Naumenkov}
\author{S.A.~Nikitin}
\author{I.B.~Nikolaev}
\author{A.P.~Onuchin}
\author{S.B.~Oreshkin}
\author{Yu.A.~Pakhotin}
\author{S.V.~Peleganchuk}
\author{S.S.~Petrosyan}
\author{V.V.~Petrov}
\author{A.O.~Poluektov}
\author{A.A.~Polunin}
\author{G.E.~Pospelov} 
\author{\fbox{\raisebox{0pt}[0.75\totalheight][0.15\totalheight]{I.Ya.~Protopopov}}}
\author{G.A.~Savinov}
\author{A.G.~Shamov}
\author{D.N.~Shatilov}
\author{A.I.~Shusharo}
\author{B.A.~Shwartz}
\author{V.A.~Sidorov}
\author{E.A.~Simonov}
\author{Yu.I.~Skovpen}
\author{A.N.~Skrinsky}
\author{A.M.~Soukharev}
\author{A.A.~Talyshev}
\author{V.A.~Tayursky}
\author{V.I.~Telnov}
\author{Yu.A.~Tikhonov}
\author{K.Yu.~Todyshev}
\author{G.M.~Tumaikin}
\author{Yu.V.~Usov}
\author{A.I.~Vorobiov}
\author{A.N.~Yushkov}
\author{A.V.~Zatsepin}
\author{V.N.~Zhilich}
\affiliation{
KEDR Collaboration. Budker Institute of Nuclear Physics, 630090 Novosibirsk, Russia
}


  \begin{abstract}
    A new high precision measurement of the \JP- and \PP-meson masses
    has been performed at the \VEPP{4M} collider using the KEDR
    detector.  The resonant depolarization method has been employed
    for the absolute calibration of the beam energy.  The following
    mass values have been obtained:
    \begin{equation*}
      \begin{aligned}
         &{\rm M}_{\JP} &= 3096.917 &\pm 0.010 \pm 0.007\ \text{MeV}, \\
         &{\rm M}_{\PP} &= 3686.111 &\pm 0.025 \pm 0.009\ \text{MeV}.
       \end{aligned}
     \end{equation*}
    The relative measurement accuracy has reached $4\cdot 10^{-6}$ for
    \JP and $7\cdot 10^{-6}$ for \PP, approximately 3 times
    better than in the previous precise experiments.  

  \end{abstract}


\maketitle

\section{Introduction}

This work continues a series of experiments on the precise
determination of the onium resonance masses at the electron-positron
collider \VEPP{4}: \JP, \PP (OLYA detector) \cite{PSI12} and $\Upsilon$,
$\Upsilon^{\prime}$, $\Upsilon^{\prime\prime}$ (MD-1 detector)
\cite{UPS1,UPS23,UPS1M,MD1}.  A few years ago the values of
the masses obtained have been rescaled \cite{ONUCHIN,REVISIT} to take
into account the progress in the electron mass
measurements~\cite{Emass1,Emass2}.

\VEPP{4} experiments employed the resonant depolarization method
\cite{BUK,DERB} for the absolute beam energy calibration and achieved
the relative mass accuracy of $1\cdot 10^{-5}$ for the
$\Upsilon$-family and of $3\cdot 10^{-5}$ for the $\psi$-family. The
resonant depolarization experiments on bottomonium masses were also
performed with the CUSB detector at CESR \cite{mCUSB} ($\Upsilon$) and
with the ARGUS detector at DORIS \cite{mARGUS} ($\Upsilon'$).  The
accuracy of the \JP-mass measurement was improved in the Fermilab
$p\bar{p}$-experiment E760 \cite{E760} to $1.2\cdot 10^{-5}$ using the
\PP mass value from Ref. \cite{PSI12}.

The goals of this work were to further improve the accuracy of the
\JP- and \PP-masses and develop the resonant depolarization technique
at the upgraded \VEPP{4M} collider for future experiments.

The first precise measurement of the $J/\psi$ and $\psi^{\prime}$
meson masses \cite{PSI12} set the mass scale in the range around 3~GeV
which provided a basis for the accurate determination of the
charmonium state location.  At present the charm meson family is a
good test bench for QCD and quark potential models predictions in
which masses of the open and hidden charm can be calculated with good
accuracy.  Another fundamental application of the mentioned
measurements is the $\tau$-lepton mass determination \cite{tauMass}.

Substantial improvement in the beam energy accuracy obtained by the
presented experiment sets a new standard of the mass scale in the
charmonium range.

\section{Beam energy determination technique}

\subsection{Resonant depolarization method\label{sec:ReDe}}

Electrons and positrons in storage rings can become polarized due to
emission of synchrotron radiation according to the Sokolov-Ter\-nov
effect \cite{SokTern}. Spins of polarized electrons precess around the
vertical guiding magnetic field with the precession frequency
$\Omega$, which in the plane orbit approximation is directly related
to the particle energy $E$ and the beam revolution frequency $\omega$:
\begin{equation}
\Omega/\omega = 1 + \gamma \cdot \mu^{\prime}/\mu_0 = 1 + \nu \:,
\label{eqn:Omega}
\end{equation}
where $\gamma=$E/m$_{\rm e}$, m$_{\rm e}$ is the electron mass,
$\mu^{\prime}$ and $\mu_0$ are the anomalous and normal parts of the
electron magnetic moment. The $\nu$ is a spin tune, which represents
the spin precession frequency in the coordinate basis related to the
particle velocity vector.

The precession frequency can be determined using the \emph{resonant
  depolarization}. To this end one needs a polarized beam in the
storage ring which is affected by the external electromagnetic field
with the frequency $\Omega_D$ given by the relation
\begin{equation}
     \Omega \pm \Omega_D = \omega \cdot n 
\label{eqn:OmegaD}
\end{equation}
with any integer $n$
(for \VEPP{4M} in the \JP region \mbox{$n=3$}).

The precession frequency is measured at the moment of the polarization
destruction detected by the \emph{polarimeter}, while the
\emph{depolarizer} frequency is being scanned.  The process of forced
depolarization is slow enough compared to the period of the
synchrotron oscillations of the particle energy.  This allows to
determine the average spin tune $\left<\nu\right>$ and corresponding
average energy of the particles $\left<E\right>$ with higher accuracy
than the beam energy spread $\sigma_E$.

Due to modulation of the precession frequency by particle orbital
motion, the resonant depolarization could happen at the sideband
resonances, which are distant from the main one by multiples of the
synchrotron and betatron frequencies. Besides, it could happen at the
weak sideband resonances caused by extraneous low frequency modulation
of the guide field, caused for example by pulsations in the power
supply system (50~Hz, the energy shift of about 25~keV). Therefore, it
is necessary to identify the main resonance by special means.

It should be noted that the average energy of the beam particles
$\left<E\right>$ differs from the energy of the equilibrium particle
$E_s$ because of the radial betatron oscillations. The effect is
proportional to the betatron amplitude squared and is mainly due to
the nonlinearity of the guide field. It also determines the spin
resonance natural bandwidth \cite{SHATUNOV}. 
In this experiment the observed full bandwidth was about
5~keV in beam energy units.

Formula (\ref{eqn:Omega}) gives the value of $\gamma$ averaged over
the beam revolution time. Thus, for a symmetric machine, it
corresponds to the energy in the interaction point.

The method described has been developed in Novosibirsk and first
applied to the $\phi$-meson mass measurement at the \VEPP{2M} storage
ring \cite{BUK}.
The comprehensive review of the resonant depolarization technique and
its applications for particle mass measurements can be found 
in~\cite{UFN}.

\subsection{\VEPP{4M} polarimeter}

The polarimeter unit is installed in the technical straight section of
\VEPP{4M} and consists of the polarimeter employing the spin
dependence of the intra-beam scattering (Touschek) effect
\cite{TOUSCHEK} and TEM wave-based depolarizer \cite{POLAR}.

The polarimeter detects Touschek electron pairs with the help of two
movable scintillation counters placed inside the beam pipe pockets.
We use the ``two bunches'' compensation technique, in which relative
rates of scattered particles from unpolarized and polarized beams are
compared.

The rate of Touschek electrons is $3\div12$~kHz at the beam current of
$2\div4$~mA. The depolarizer frequency is scanned with a step of
$2$~Hz by the computer controlled synthesizer with the intrinsic
bandwidth of $\sim1$~Hz. However, the frequency line is artificially
broadened up to the $4$~Hz band. This provides controllable conditions
\cite{CHICAGO} for the depolarization at the main spin resonance at
the minimal level of the depolarizer power, which corresponds to
$\sim2\cdot10^{-6}$~rad spin rotation per a single pass of the
particle, and with the frequency-tuning rate of $0.2$~Hz/sec.

\begin{figure}[htb]
\centering\includegraphics[clip,width=\columnwidth]{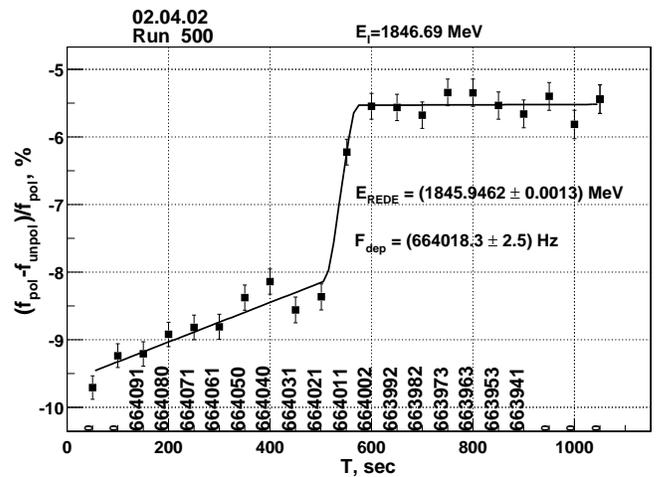}
\caption{\label{fig:Jump} The variation of the coincidence
  rate ratio for polarized and unpolarized beams
during the energy calibration ($E_I$ is the energy calculated from the
magnet currents, $E_{REDE}$ is the energy determined by resonant 
depolarization,
$F_{dep}$ is the corresponding depolarizer frequency;
the vertical numbers show
the instant depolarizer frequency).
}
\end{figure}

The characteristic jump in the relative rate of scattered electrons at
the moment of resonant depolarization is $3.0\div3.5$\% with the
statistical error of $0.3\div0.4$\% for the beam polarization degree
higher than 50\%.  Typical behavior of the rate ratio is shown in
Fig.~\ref{fig:Jump}.  The linear growth before the depolarization
reflects the difference in the bunch life times due to polarization
dependence of the intra-beam scattering cross section.

The characteristic uncertainty of the beam energy calibration due to
the depolarization procedure is 1.5~keV (for more detail see Ref.
\cite{POLAR}).

\subsection{Accuracy of single energy calibration\label{sec:Cerr}}

The achievable accuracy of the \JP and \PP mass measurement was
analyzed in \cite{BOG}. Since that time, understanding of some
systematic effects improved. In particular, the energy shift due to
the \emph{vertical closed orbit disturbances} turned out to be much
less than was expected in \cite{CHICAGO} and \cite{BOG}.

The relation (\ref{eqn:Omega}) is broken in the radial magnetic field
and the vertical electric field (used for the electrostatic beam
separation) because of the orbit non-planarity.  The influence of
these fields in the second order of perturbation theory can be
expressed in terms of Fourier harmonics of the vertical closed orbit
disturbances.  The accurate analysis of the effect and the numerical
simulation gave the energy shift of \mbox{$-0.4\pm0.3$~keV} for \JP
and \mbox{$-0.3\pm0.2$~keV} for \PP.

The uncertainty estimates for the \emph{mean value of the beam energy
  in the interaction point} for a single calibration are collected in
Table~\ref{tab:ErrC}.

\begin{table*}[ht]
\caption{\label{tab:ErrC}Single energy calibration uncertainties 
in the vicinity of \JP and \PP~(keV). }
\begin{tabular}{p{0.48\textwidth}p{0.3\textwidth}ll}\hline
\emph{Source}  & \emph{Nature} & \JP & \PP \\
\hline
Vertical orbit disturbances       &Systematic 
       & 0.3$^{*}$ & 0.2$^{*}$ \\
Spin resonance width              &
    Systematic, depends on the frequency scan direction  & 
                               1.0$^{*}$ & 1.0$^{*}$ \\
Coherent energy loss asymmetry    &Systematic, charge depending& 0.6 & 1. \\
Precession frequency measurement accuracy    &Statistical  & 1.2    & 1.5  \\ 
Revolution frequency measurement accuracy of $10^{-8}$
                                             &Statistical & 0.2    & 0.2  \\
\hline
\emph{Sum in quadrature}                        &      &  1.7    &  2.1  \\ 
\hline
\multicolumn{4}{l}{$^{*}$~--- correction uncertainty}
\end{tabular}
\end{table*}

The energy value obtained in a single calibration is biased due
to a \emph{non-zero spin resonance width}.
The required correction ($\approx 2.5$~keV) can be
determined with accuracy better than 1~keV using
a few calibrations with opposite directions of the depolarizer
frequency scan.

The \emph{coherent energy loss asymmetry}\, arises from the difference
of the resistive impedance of the beam pipe in two half-rings.
As a result, the energy of electrons and positrons
in the interaction point (I.P.)
differs from the energy value obtained by the resonant depolarization.

The uncertainties not exceeding 0.1~keV are not shown in the table
including those due to the \emph{non-zero momentum compaction factor}
and the \emph{longitudinal magnetic fields} \cite{BOG}.

The \emph{uncertainty of a single energy measurement} does not 
directly contribute
to the \emph{systematic error of the meson mass}. 
Thus, the effect of
the \emph{energy loss asymmetry}
in the half-rings has an opposite sign for $e^{+}$
and $e^{-}$ and cancels in the linear approximation. 
The contribution of the \emph{precession and revolution frequency
measurements} has mainly statistical nature and
becomes negligible when a large number of calibrations is used.
At the same conditions the uncertainty due to the \emph{non-zero spin resonance
width} vanishes provided that frequency scan directions alternate.

On the other hand, new sources of the systematic error come into play
when a long-term experiment with colliding beams is considered. This
is discussed below in Sec.~\ref{sec:Eass}, Sec.~\ref{sec:MeanW} and,
finally, Sec.~\ref{sec:ErrDisc}, when the essential features of the
experiment are described.

\section{Experiment description}\label{sec:Exp}

The first part of the experiment consisted of three scans of the
\JP-region (the integrated luminosity  
$\int\! L dt \approx 40$~nb$^{-1}$, 
the beam energy spread $\sigma_E \approx 0.6$~MeV) and
three scans of the \PP-region 
($\int\! L dt \approx 76$~nb$^{-1}$, $\sigma_E \approx 0.9$~MeV).
Then the betatron and synchrotron damping decrements of \VEPP{4M}
were rearranged to reduce the energy spread to 0.45~MeV and
the fourth scan of \JP was performed ($\int\! L dt \approx 10$~nb$^{-1}$).
The goal was to verify possible systematic errors related
to the collider operating mode and the beam energy spread.

The beam polarization time in the \VEPP{4M} ring is about 100 hours at
the $\psi$-energy region. For the energy calibration runs, the beam spent
the time sufficient for the polarization in the booster ring \VEPP{3}
(2.5 hours at \JP and about 1 hour at \PP) and was injected to \VEPP{4M}
 without essential loss of the polarization degree.

\begin{figure}[tb]
\includegraphics[clip,width=\columnwidth]{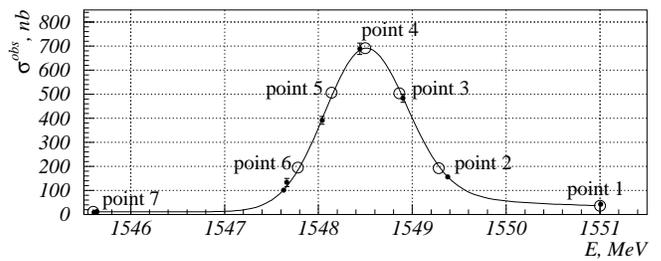}
\caption{\label{fig:aScan} The data acquisition scenario for \JP
  (circles) and the actual points of the second scan (points with
  error bars).  The solid line shows the second scan fit.  }
\end{figure}

During the scan the data were collected at 7 points of the
resonance excitation curve (Fig.~\ref{fig:aScan}). At points
1 and 7 the required integrated luminosity was reduced by the 
factor of 2. Such a $5+2$ scheme does not minimize statistical  
and systematic errors for a given experiment duration
but allows one to apply the $\chi^2$ criterion
to the results of a single scan.

Before data acquisition at point 1,
the beam energy calibration was made to fix the current energy scale.
At points 2--6 the calibrations before and after data taking were
performed with the opposite direction of the depolarizer frequency
scan. The point 7 requires no energy calibration. From 2 to 5 ring
fillings were necessary at each point to collect the wanted
integrated luminosity.
The injection from \VEPP{3}
occurred at the required set-up energy without any intentional
change of the \VEPP{4M} magnet currents.

On completion of the scan, the \VEPP{4M} magnetization cycle was
performed and the whole procedure was repeated. 

The set-up parameters of the collider and the results of the current,
magnetic field,
temperature and orbit measurements have been stored in the database.

\section{Assignment of energy to data acquisition runs}

Unlike the experiments \cite{UPS1,UPS23,UPS1M}, the energy
calibration during the data acquisition was not possible, therefore it 
was necessary to assign the energy to each data acquisition run
using the results of the energy calibration runs.

\begin{figure}[tb]
\begin{center}
\includegraphics[clip,width=\columnwidth]{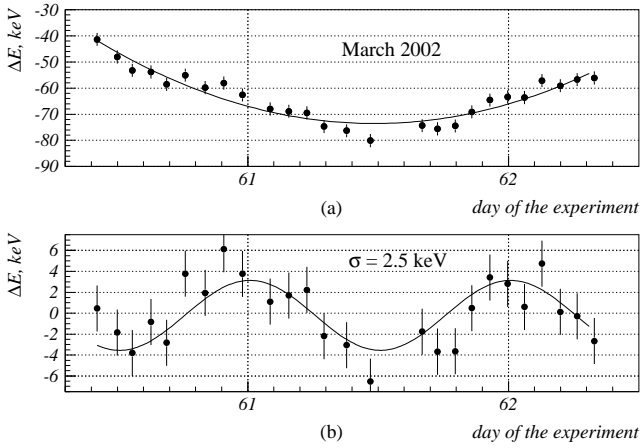}
\end{center}
\caption{\label{fig:Stab1} The results of the first stability run: (a)~---
parabolic fit of the energy deviation, (b)~--- the same with the fit
subtracted
(the error bars show the mean deviation from this fit).
}
\end{figure}

\begin{figure}[tb]
\begin{center}
\includegraphics[clip,width=\columnwidth]{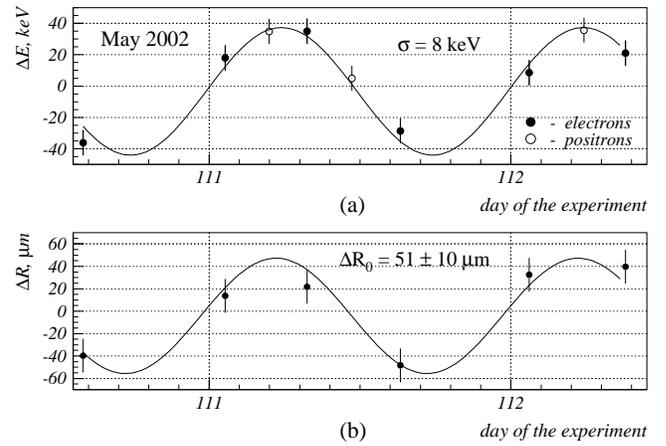}
\end{center}
\caption{\label{fig:Stab2} The results of the second stability run: (a)~---
energy deviation,
(b)~--- the average orbit position given by BPMs where available
(the error bars show the mean deviation from the fit).}
\end{figure}

\subsection{Stability runs and energy prediction function}\label{sec:Stab}

To make the reliable energy assignment, two \emph{stability runs}
consisting of the continuous series of energy calibrations without
any intentional magnet current changes were
performed, one after the third scan of \PP (March 2002) at
the beam energy $E \approx 1846$~MeV \cite{STAB}
and the other one after the fourth scan of \JP (May 2002) at
$E \approx 1550.6$~MeV.
Their results are presented in Figs.~\ref{fig:Stab1},\ref{fig:Stab2}.

One can see a rather slow energy variation and the day-to-night 
oscillation with the amplitude growing from 4~keV in March to 45~keV 
in May. The energy oscillations with the amplitude of \mbox{($45\pm5$)}~keV
observed since the 110th day of the experiment
are correlated with time dependence
of the average orbit position with the amplitude of 
\mbox{$\Delta R = (51\pm10)$}~$\mu$m. The estimate of the corresponding 
energy deviation is \mbox{($77\pm15$)}~keV under the assumption that 
the collider expands and shrinks uniformly.

The large and relatively fast energy variations do not allow us to use
the mean energy of the two
calibrations surrounding a data acquisition run as the beam energy
for this run, as was supposed initially. Instead,  \emph{energy
prediction functions} have been suggested, which employ
the results of the
field measurements in some magnets by nuclear magnetic resonance (NMR)
and the temperature measurements and include the explicit 
time dependence as a substitute of variables, which were not monitored
(the effect of the tunnel wall temperature on the ring perimeter etc.).
The orbit measurements
at \VEPP{4M} are not accurate and comprehensive enough to be used
for the energy prediction (there are ten independent bending magnet
power supplies, sixty radial correctors and only fifty four 
beam position monitors).

Several prediction functions have been tried of the form:
\begin{equation}
\begin{split}
 &E_p = \mathscr{P}\cdot H_{NMR}\cdot
       ( 1+\varkappa\cdot( T_{ring} - T_{NMR}))\times \\
                        &\hspace{3.9cm}f(T_{ring},T_{air},T_{water}) +\\
       &\delta E_{on} \cdot \exp\left(-\frac{t_{on}}{\tau_{on}}\right) +
 \delta E_{cycle} \cdot \exp\left(-\frac{t_{cycle}}{\tau_{cycle}}\right) +\\
       &A(t)\cdot \cos\left(\frac{2\pi t}{\tau}-\varphi(t)\right) + 
  E_0(\Delta i,t),
\end{split}
\end{equation}
where $H_{NMR}$ is the field in the off-ring calibration magnet with
the temperature of $T_{NMR}$; $T_{ring}$, $T_{air}$ and $T_{water}$
are the average values of the ring and cooling agent temperatures, $t$
is current time, $t_{on}$ and $t_{cycle}$ denote time elapsed since
the last switching on the collider and the last magnetization cycle,
respectively. The $\mathscr{P}$, $\varkappa$, $\delta\!E_{on}$,
$\delta\!E_{cycle}$, $\tau_{on}$, $\tau_{cycle}$ and $\tau$ are free
parameters determined by the fit of all energy calibrations performed
in a certain operation mode (\JP scans I$\div$III, \PP, \JP scan IV).

\begin{figure}[t]
\begin{center}
\includegraphics[clip,width=\columnwidth]{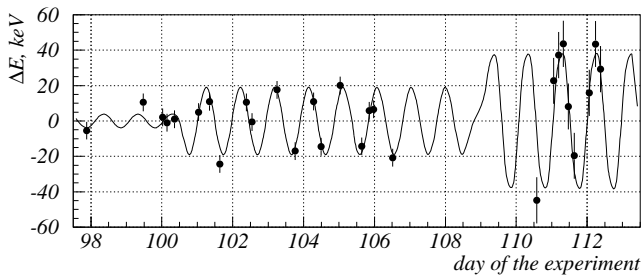}
\end{center}
\caption{\label{fig:Sin} The predicted energy with all aperiodic
dependences removed (the error
bars show the mean deviation from the prediction).}
\end{figure}

The first term is the basic parameterization of the magnetic
field integral behavior.
The exponential terms introduce the magnetic field relaxation;
ignoring this effect results in a bias in the parameters
$\mathscr{P}$, $\varkappa$ etc.
The oscillating term describes the variations of the ring perimeter
under the assumption that the Fourier harmonics around $\Omega=2\pi/\tau$
dominate. The $A$ and $\varphi$ are not constant due to season
variations in properties of the ground surrounding the machine
tunnel. The term $E_0(\Delta i,t)$ takes into account the energy variation due
to a few relatively large adjustments of the current $\Delta i$ 
in some machine lattice elements.

The prediction functions differ by the choice of
$f(T)$, $A(t)$ and $\varphi(t)$. These (simplest) functions, as well
as $E_0(\Delta i,t)$, have additional free parameters. 

\begin{figure}[t]
\begin{center}
\includegraphics[clip,width=\columnwidth]{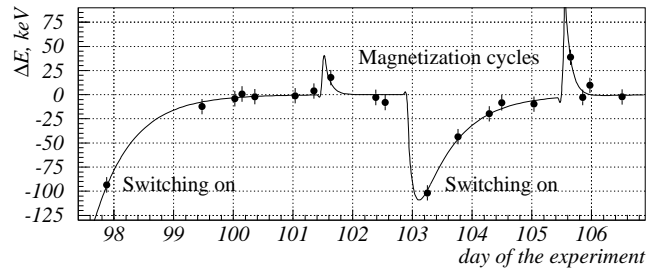}
\end{center}
\caption{\label{fig:On} Aperiodic energy dependence on time due to
switching on the ring and the magnetization cycles (the error
bars show the mean deviation from the predicted energy).}
\end{figure}

The results of the best energy prediction for May 2002 (the fourth
scan of \JP and the second stability run) are illustrated by
Figs.~\ref{fig:Sin} and \ref{fig:On}.  It should be noted that the
time dependence presented in Fig.~\ref{fig:On} is partially
compensated by the temperature and field strength dependences.
Fourteen free parameters were used to fit 28 points shown in these
figures.  The value of the chi-squared per degree of freedom was
employed to estimate the (mean) error of the energy prediction by
requiring $\chi^2/N_{DoF}=1$.  According to the fit, the energy
oscillation period \mbox{$\tau = 1.02\pm 0.02$}~days, so we have fixed
it at the value of 1.

The appearance of strong oscillations (Fig.\ref{fig:Sin}, the 100th
day of the experiment) and their further growth (between 107th and
110th days) can be probably explained by the change of the
thermomechanical properties of the ground surrounding the half-ring
tunnels. These properties can change abruptly at the moment when the
melting front reaches the tunnel (in Novosibirsk it occurs in May).
According to the $\chi^2$ criterion, the sudden growth of the
oscillation amplitude on the 100th day is much more probable than the
gradual one.

The direction of the depolarizer frequency scan was taken into account
in the prediction function fit. This allows us to determine the spin
resonance half-width of \mbox{$2.4 \pm 0.7$}~keV in beam energy units
and makes the prediction function value unbiased.  The accuracy of the
energy prediction varies from 6 to 8~keV during the whole experiment
(218 calibrations).

\subsection{Energy assignment accuracy}\label{sec:Eass}

The uncertainty of the energy prediction of $(6 \div 8)$~keV includes
the \emph{statistical error of the single energy calibration} and all
the \emph{uncertainties due to the run-to-run difference}. Among the
sources of the latter, the \emph{radial closed orbit variations} and
the \emph{RF frequency drift} dominate \cite{BOG}. The model of
statistically independent orbit perturbations used in \cite{BOG}
overestimates the effect of the radial closed orbit variations.

The significant difference between the energy calibration runs and the
data acquisition runs was the beam separation at the additional
interaction point. This separation could cause some small
sign-of-charge dependent shift of the $e^{+}$- and $e^{-}$-beam
energy.  \emph{During the calibrations, the separation was turned off},
so in the linear approximation the energy of the single beam is equal
to the mean energy of two beams with the separation on.  \emph{During
  the data acquisition, the separation was turned on} to provide
normal conditions for beam-beam effects.  This increases the length of
the beam orbit in the straight section and decreases it in the
half-rings causing the equal energy shift for the $e^{+}$- and
$e^{-}$-beams. The estimated \emph{energy shifts are $(-1.7\pm0.2)$~keV
  at \JP and $(-2.0\pm0.2)$~keV at \PP}.

Due to the collision effects, the amplitude of the radial betatron
oscillations in the data acquisition runs is bigger than that during
the energy calibration, therefore, the average particle energies are
not the same (see Sec.~\ref{sec:ReDe}).  The corresponding systematic
errors of the energy assignment do not exceed 1.5~keV for \JP and
1.8~keV for \PP.

With the two exceptions mentioned above, the energy calibration runs
and the data acquisition runs do not differ essentially and we assume
the same \emph{statistical accuracy} of the energy prediction for them.

\begin{table*}[t]
\newcommand{\LT}{$<\,$}
\newcommand{\AP}{$\approx\,$}
\newcommand{\BL}{\phantom{$\approx\,$}}
\caption{\label{tab:ErrR}
The energy uncertainty for the data acquisition runs  
in the vicinity of \JP and \PP~(keV).}
\begin{tabular}{p{0.4\textwidth}p{0.3\textwidth}ll}
\hline
\emph{Source} & \emph{Nature}  &\BL\JP & \BL\PP \\
\hline
Energy prediction                      & Statistical     &\BL 7.6 & \BL 6.5 \\ 
Radial betatron oscillations           & Systematic      &\LT 0.7  &\LT 0.9 \\ 
Single energy calibration (Sec.~\ref{sec:Cerr}) 
                                       & Systematic      &\BL 0.3  &\BL 0.2 \\ 
                                       & Charge depending &\BL 0.6  &\BL 1.0 \\
Beam separation in the additional I.P.
                                      & Systematic
                                              &\BL 0.2$^{*}$ &\BL 0.2$^{*}$ \\ 
\hline
\emph{Sum in quadrature}                &               &\AP 7.7  &\AP 6.6 \\ 
\hline
\multicolumn{4}{l}{$^{*}$~--- correction uncertainty}
\end{tabular}
\end{table*}

Estimates of the beam energy uncertainty for the data acquisition runs
are presented in Table~\ref{tab:ErrR}.  The statistical component of
the beam energy uncertainty contributes to the meson mass error with
the factor $\propto 1/\sqrt{N}$, where $N$ is the number of
calibrations. The accurate calculation of this contribution as well as
that due to the choice of the energy prediction function can be
performed using the actual resonance curve fit (see
Sec.~\ref{sec:ErrDisc}).

\section{Determination of mean collision energy}\label{sec:MeanW}

The production rate of the resonance of mass $M$ at the given collider
energy $E$ is determined by the probability of $e^{+}e^{-}$-collisions
with the invariant mass \mbox{$W \simeq M$}.  Taking the angular
spreads $\theta_x$, $\theta_y$ and the energy spread $\sigma_E$ into
account, one has after averaging over the particle momenta
\begin{equation}
  \begin{split}
    \left<W\right>_{p} \approx
    \left<E_{+}+E_{-}\right> &-
    \frac{1}{2} (\theta_x^2+\theta_y^2) E -\\
    &\frac{\sigma_E^2}{2E} -
    \frac{(\left<E_{+}\right>-\left<E_{-}\right>)^2}{4E}.
  \end{split}
\label{eqn:Wmean}
\end{equation}
The value of $E$ is determined by the resonant depolarization; the
last term is due to the difference of the coherent energy loss in two
half-rings (Sec.~\ref{sec:Cerr}).  For the \VEPP{4M} conditions at the
$\psi$-energy region the correction is about 0.2~keV, thus we assume
$W=E_{+}\!+\!E_{-}$ for each collision.

For beams with the Gaussian energy spread in the presence of the
electrostatically induced vertical dispersion $\psi^{*}_y$ and the
beam impact parameter $\Delta_y$, the differential luminosity can be
written as
\begin{equation}
\begin{split}
 &\frac{d\Lum(E,W)}{dW} = 
 \frac{f_R N_{+}N_{-}}{4\pi\sigma^{*}_x(W/2)\sigma^{*}_y(W/2)} \cdot
  \frac{1}{\sqrt{2\pi}\,\sigma_W} \times\\
  &\exp{\left\{-\frac{1}{2} \left(
   \frac{W-2E}{\sigma_W}-\frac{\sigma_W\psi^{*}_y\Delta_y}{2E\sigma_y^2}
  \right)^{2}-\frac{\Delta_y^2}{4\sigma_y^2}\right\}} ,
\label{eqn:L}
\end{split}
\end{equation}
where $f_R$ is a revolution frequency, $N_{+}$ and $N_{-}$ are the
bunch populations.  The transverse beam sizes in the interaction point
$\sigma^{*}_x$, $\sigma^{*}_y$ effectively depend on the sum
\mbox{$E_{+}\!+\!E_{-}$} due to the collider \mbox{$\beta$-function}
chromaticity (the formula is valid to the first order of this effect
and assumes the beam symmetry).  According to (\ref{eqn:L}), the mean
collision energy \mbox{$\left< W\right>_L = \int\!w d\Lum(E,w) \ne
  2E$}, which leads to the systematic error in the resonance mass.

The vertical dispersion \mbox{$\vert\psi^{*}_y\vert \approx 800$}~$\mu$m with
opposite signs for $e^{+}$ and $e^{-}$ appears in \VEPP{4M} due to the
beam separation in the additional interaction point.  The residual
orbit perturbations related to this separation result in the beam
misalignment in the experimental I.P. characterized by $\Delta_y$.
It leads
simultaneously to the luminosity loss and the collision energy shift.
To avoid that, the voltage on the experimental I.P. separator plates
was tuned to have maximum luminosity in each run with the accuracy
better than 3\% so that \mbox{$\Delta_y < 2.5$}~$\mu$m  at the beam size
\mbox{$\sigma^{*}_y \simeq 7$}~$\mu$m.  Thus, the uncertainty of the
collision energy $W$ in the run is less than 10~keV for \JP and less
than 18~keV for \PP.  In the resonance mass error, this uncertainty is
suppressed $\propto 1/\sqrt{N}$, where \mbox{$N > 100$} is the number
of runs (see Sec.~\ref{sec:ErrDisc}).

The $\beta$-function chromaticity also leads to the shift of the mean
value of the collision energy.  For the given emittance $\epsilon_y$,
the vertical beam size \mbox{$\sigma^{*}_y =
  \sqrt{\epsilon_y\beta^{*}_y}$}.  Using approximations
\mbox{$\beta_y(W/2)\approx\beta^{*}_y\,
  (1\!+\!\partial\ln\beta^{*}_y/\partial E \,\,(W/2\!-\!E))$} and
\begin{equation}
\begin{split}
 &\frac{1}{\sqrt{1+\partial\ln\beta^{*}_y/\partial E\,\,(W/2\!-\!E)}}
   \approx\\
  &\hspace{1cm}\exp{\left\{
  -\frac{1}{2}\partial\ln\beta^{*}_y/\partial E\,\,(W/2\!-\!E)
  \right\}}
\end{split}
\end{equation}
in (\ref{eqn:L}), one obtains
\mbox{$\delta\left<W\right>=-1/4\,\,
\partial\ln\beta^{*}_y/\partial E\,\,\sigma_W^2$}.
The effect of the radial $\beta$-function chromaticity
is suppressed by the factor of
\mbox{$(\sigma^{*}_{x,\beta}/\sigma^{*}_x)^2 \sim 0.1$} where
$\sigma^{*}_{x,\beta}$ is the betatron radial size and
$\sigma^{*}_x$ is the total size including the dispersion.
The $\beta^{*}(E)$
 measurements at \VEPP{4M} gave the shifts of
\mbox{$-4 \pm 2$}~keV for \JP scans I$\div$III,
\mbox{$-1.5 \pm 0.7$}~keV for \JP scan IV
and \mbox{$+5 \pm 2.5$}~keV for \PP (the chromaticity was
partially compensated after the third scan of \JP).

At the 1~keV level of accuracy the potential energy of colliding
particles inside the beams should be taken into account.  The
effective energy of the electron is \mbox{$E_{kinetic}+U/2$}, where
the potential energy $U$ is due to its Coulomb interaction with all
other electrons of the beam.  For the flat beam with the logarithmic
accuracy
\begin{equation}
 U = \frac{e^2N}{\sqrt{\pi}\,\sigma_z} \ln{\frac{D}{\sigma_x}},
\end{equation}
where $N$ is the bunch population, $\sigma_z$ is the longitudinal
bunch size and $D$ is the beam pipe diameter (in the beam rest frame
the interaction of particles at longer distances is screened out). The
kinetic and potential energies in the I.P. differ from those in the
ring because of the difference in the beam and beam pipe sizes, but
the total energy conserves during the revolution, therefore
\begin{equation}
  E_{kinetic, I.P.}+U_{I.P.}/2
  = 
  E_{kinetic, ring}+U_{ring}/2
\end{equation}
At the moment of the annihilation the total energy of the $e^{+}e^{-}$
pair transforms to the product mass, thus
\begin{equation}
  \begin{split}
    W = &2\cdot (E_{kinetic, I.P.}+U_{I.P.}) =\\
    &\hspace{1cm}2E_{kinetic, ring}+U_{ring}+U_{I.P.}
\end{split} 
\end{equation}
The resonant depolarization result is 
\mbox{$\approx E_{kinetic, ring}$},
therefore the collision energy shift 
\mbox{$\delta W=U_{I.P.}+U_{ring}$}.
The energy losses ignored in this consideration does not
change the final result.

For the the actual values of the beam currents and sizes it leads
to a correction of \mbox{($2\pm1$)}~keV for
\JP and \PP.

\section{Event selection and luminosity measurements}
\subsection{Detector and trigger}

The KEDR detector \cite{KEDR} consists of the vertex detector, the
drift chamber, the time-of-flight system of scintillation counters,
the particle identification system based on the aerogel Cherenkov
counters, the calorimeter (the liquid krypton in the barrel part and
the CsI crystals in the end caps) and the muon tube system inside and
outside of the magnet yoke.  In this experiment the magnetic field was
off and the liquid krypton calorimeter was out of operation.

To suppress the machine background to the acceptable level, the
following trigger conditions were used by OR
\begin{enumerate}
\item
     signals from $\ge$ 2 barrel scintillation counters coinciding with the
                       CsI calorimeter signal,
\item coinciding signals of two CsI end-caps,
\end{enumerate}
with the CsI energy threshold of about 75~MeV.  The Monte Carlo
simulation employing the JETSET-7.4 code \cite{JETSET} yields the
trigger efficiency of about 0.4 for \JP decays and about 0.43 for \PP
decays.
\subsection{Multihadronic event selection}\label{sec:EvSel}
For the off-line event selection the following conditions were
applied:
\begin{enumerate}
\item $\ge$ 3 charged tracks or 2 acolinear charged tracks
 ($\cos\theta<0.95$)
 from the interaction region (\mbox{$\rho<5$}~mm, \mbox{$|z|<120$}~mm),
\item the CsI energy $>$ 1.15 of the hardware threshold.
\end{enumerate}
The second condition serves to exclude the hardware threshold
instability.  The detection efficiency, determined by the visible peak
height and the table value of the leptonic width, is about 0.25 for
\JP (\mbox{$\sim 20\cdot 10^{3}$} events) and about 0.28 for \PP
(\mbox{$\sim 6\cdot 10^{3}$} events).

The residual machine background (beam-gas and beam-wall) does not
exceed \mbox{5 nb}.  The systematic error in the meson masses related
to its variation is less than 1~keV. To obtain this estimate, the
background was increased by a few times by adding the appropriate
fraction of unselected events to the selected ones at each
experimental point (see Fig.~\ref{fig:aScan}). Further suppression of
the background leads to the detection efficiency loss and does not
improve the mass accuracy.

The meson mass value is rather sensitive to the detection efficiency
variation during the energy scan, its reduction by 1\% at one point
causes the \PP mass shift up to 5~keV. To ensure the detection
efficiency stability, all electronic channels having problems at any
moment of the experiment were \emph{excluded from the off-line
  analysis}. Besides, the relative \emph{hit efficiencies} of all
detector subsystems were obtained for all experimental points using
the cosmic ray runs, the multihadron event statistics and (when
possible) background events.  These efficiencies were applied to
\emph{the real multihadron events} to determine the relative
point-specific correction factors. The variation of the drift chamber
spatial resolution was handled similarly.

The correction procedure described above shifts the mass by
\mbox{($+6.3\pm2.3$)}~keV for \JP and \mbox{($+0.2\pm2.0$)}~keV for
\PP.  The errors include statistical errors of the hit efficiency
determination and the uncertainties of the correction procedure
employed. The shift of the \JP mass is mainly due to the false alarm
of the safety system which stopped the gas flow in the drift chamber
(one point of the second scan). The values of shifts are given just
for information, only the errors are of importance.

\subsection{Luminosity measurements}\label{sec:LumMeas}
For the operative \VEPP{4M} luminosity measurements single
bremsstrahlung monitors were installed in both $e^{+}$- and
$e^{-}$-directions.  Their stability is not sufficient for the
precision mass measurements, therefore Bhabha-scattering events
detected by the end-cap CsI calorimeter were employed. The fiducial
polar angle interval is \mbox{$17.5^{\circ}<\theta<35^{\circ}$}.  The
resonance contribution is not negligible in this angular range, so
that the $\ee \to \psi \to \ee$ and $\ee \to \ee$ interference should
be taken into account.  The correction to the number of \ee-events has
been included according to \cite{AZIMOV}. The values of the total and
electronic widths were taken from \cite{PDG} and the beam energy
spreads were known from the experiment.

Unlike $\ee \to \mumu$, the $\ee \to \ee$ interference dip is at the
high-energy side of the resonance curve, so the mass shifts due to the
correction are positive: \mbox{($15 \pm 1$)}~keV~(\JP) and \mbox{($5
  \pm 0.5$)}~keV~(\PP).

To the errors quoted, 2~keV should be added for \JP and 3~keV for \PP
to cover the calorimeter instabilities.

\section{Resonance excitation curve fitting}

\subsection{Introduction}\label{sec:FitIntro}

Taking into account the beam energy spread, the event production rate
for the collider energy $E$ can be written as
\begin{equation}
   F(E) = \int\! \sigma(W)\: d\Lum(E,W)\,,
\label{eqn:f}
\end{equation}
where $W$ is the c.m. energy of the collision, $d\Lum(E,W)$ is the
differential luminosity and $\sigma(W)$ is the cross section.

In case of the narrow vector meson production in the reaction
$e^{+}e^{-} \to V \to hadrons$
\begin{equation}
\begin{split}
  \sigma(W) &= \frac{3\pi}{M^2} \int dx\times \\
&\frac{\Gamma_{ee}\Gamma_h}{(W(1-x)-M)^2+\Gamma^2/4}
\mathscr{F}(x,W),
\end{split}
\label{eqn:KF}
\end{equation}
where $\Gamma$, $\Gamma_{ee}$ and $\Gamma_h$ are total and partial
widths of the meson, $M$ is its mass and $\mathscr{F}(x,W)$ is the
probability to lose the fraction of energy $x$ because of the initial
state radiation \cite{KF} (we substituted the Breit-Wigner cross
section with the physical value of $\Gamma_{ee}$ including the vacuum
polarization effects and used $W$ instead of $s=W^2$).

After the corrections introduced in Sec.~\ref{sec:MeanW} to exclude
the asymmetry causing the resonance mass shift, the symmetric
expression for the differential luminosity can be used:
\begin{equation}
\begin{split}
  \frac{d\Lum(E,W)}{dW}& \approx 
  \frac{L\,\left(1+k\,(W\!-\!2E)^2\right)}{\sqrt{2\pi}\,\sigma_W(I,\mathcal{J})}
\times\\
  &\hspace{1.5cm}
  \exp{\left\{-\frac{(W\!-\!2E)^2}{2\,\sigma_W^2(I,\mathcal{J})}\right\}},
\label{eqn:Lunbiased}
\end{split}
\end{equation}
where the (free) small parameter $k$ is introduced to cover
non-gaussian effects due to the $\beta$-function chromaticity and
other possible reasons.  The collision energy spread $\sigma_W$ can
depend on the beam current $I$ because of the microwave instability
reviewed in \cite{SHAP} and/or on the current density $\mathcal{J}$
due to the multiple intra-beam scattering (\cite{URAKAWA} and
references therein). The latter depends not only on the beam current,
but also on the beam sizes modifying substantially by the collision
effects, thus it must be considered as an independent parameter.

Formula (\ref{eqn:KF}) ignores the interference between the resonant
and nonresonant hadron production.  With the sufficient accuracy
\cite{AZIMOV}:
\begin{equation}
\begin{split}
\sigma(W)  = &\frac{12\pi}{M^2} \Big\{
\left(1+\frac{3}{4}\beta \right) \frac{\Gamma_{ee}\Gamma_h}{\Gamma M}
\cdot \mathrm{Im} f -\\ 
& \left(1+\frac{11}{12}\,\beta \right)
 \frac{2\alpha\sqrt{R\Gamma_{ee}\Gamma_h}}{3M}  \lambda
\cdot \mathrm{Re} f 
\Big\}.
\label{eqn:AZIMOV}
\end{split}
\end{equation}
Here $\alpha$ is the fine structure constant,
\mbox{$R=\sigma^{(h)}/\sigma^{(\mu\mu)}$}, $\lambda$~denotes \emph{the
  fraction of events interfering with the nonresonant hadronic cross
  section} and
\begin{equation}
\beta = \frac{4\alpha}{\pi} \left( \ln\frac{W}{m_e} -
\frac{1}{2}\right)
,\ 
  f  = \left( \frac{M/2}{-W+M-i\Gamma/2}
      \right)^{1-\beta}.
\nonumber
\end{equation}

In the limit of zero resonance width \mbox{$\Gamma\to0$}, assuming the
Gaussian energy spread and ignoring the interference, it is possible
to express (\ref{eqn:f}) in terms of known functions:
\begin{equation}
\begin{split}
  F(E) = &\frac{6\pi^2}{M^2}\,\frac{\Gamma_{ee}\Gamma_h}{\Gamma}
         \left(\frac{2\sigma_W}{M}\right)^{\!\beta}\times\\
         &\frac{\Gamma(1+\beta)}{\sqrt{2\pi}\,\sigma_W}\,
         \exp{\left\{-\frac{(W\!-\!2E)^2}{4\sigma_W^2}\right\}}\times\\
         &\hspace{1cm}\:D_{-\beta}\left(-\frac{W\!-\!2E}{\sigma_W}\right)
         (1+\delta)\: L,
\label{eqn:Dbeta}
\end{split}
\end{equation}
where $\Gamma$ is the gamma-function, $D_{-\beta}$ is the Weber
parabolic cylinder function and
\begin{equation}
   \delta = \frac{\alpha}{\pi}\left(\frac{\pi^2}{3}-\frac{1}{2}\right)
   + \frac{3}{4}\,\beta.
\nonumber
\end{equation}

\subsection{Interference effect treatment}\label{sec:Interf}

In principle, the interference magnitude can be left free in the fit
and extracted from the data together with the resonance mass and the
machine energy spread $\sigma_W$. Unfortunately, this affects too much
the statistical accuracy of the mass measurements, so one has to fix
it. The zero magnitude is usually assumed.

In this analysis we fix the interference parameter $\lambda$,
employing the parton model of the onium decays.  It assumes that \JP
decays to the light $q\bar{q}$-pairs with the probability of $R
B_{\mu\mu}$ and to the gluon triplet $ggg$ or the $gg\gamma$ mixture
with the probability of $1-(R+2)B_{\mu\mu}$, where $B_{\mu\mu}$ is the
muon branching ratio. The events $\JP \to q\bar{q}$ are identical to
those in the non-resonant continuum, and there is the 100\%
interference in this case.  For the hypothetical heavy onium decays,
$ggg$-$q\bar{q}$ interference is negligible due to the difference in
the angular distributions (three jets vs two jets). For the real \JP,
the angular distributions do not differ much but the interference
phases are individual for all exclusive final states, therefore the
net interference effect should be small just because of the large
number of decay modes.

So, the fraction of multihadron \JP decay events interfering with the
continuum $\lambda \approx R B_{\mu\mu}/(1-2 B_{\mu\mu}) \approx
0.17$. The uncertainty of $\lambda$ related to the final number of the
decay modes was estimated by the multiple assignment of the arbitrary
interference phases to all \JP and \PP decay modes implemented in the
JETSET-7.4 Monte Carlo code \cite{JETSET}.  The following values have
been obtained: \mbox{$\lambda_{\JP}=0.17 \pm 0.03$} and
\mbox{$\lambda_{\PP}=0.023 \pm 0.009$}.  The corresponding mass shifts
are \mbox{($+7.0 \pm 1.3$)}~keV and \mbox{($+2.0 \pm 0.8$)}~keV,
respectively. The accuracy of the parton model predictions used gives
a small addition to the errors quoted.  The fitting procedure
automatically shifts the mass value so only the errors of the quoted
values are of importance.

\begin{figure}[tb]
\begin{center}
\includegraphics[clip,width=\columnwidth]{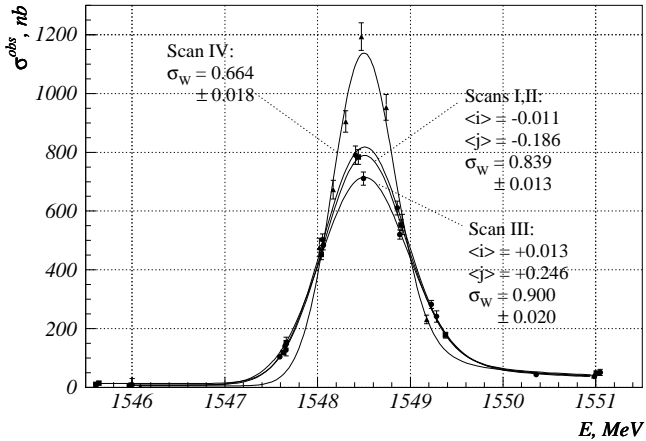}
\end{center}
\caption{\label{fig:Jpsi} The results of four \JP
  scans (the energy spread values $\sigma_W$ and the mean $i$, $j$
  values for the first two and the third scans are presented).  }
\end{figure}

\subsection{On collision energy spread variations}\label{sec:SigmaW}

The single beam measurements of the longitudinal and radial bunch
sizes indicate that the energy spread in \VEPP{4M} depends on the bunch
current.  In the \JP region the dimensionless slope
\mbox{($d\sigma_E/\sigma_E)/(dI/\left<I\right>) \simeq 0.07$}.  As the
current decreases during the run, it leads to the symmetric distortion
of the collision energy distribution and contributes to the $k$
parameter of (\ref{eqn:Lunbiased}). If the mean value of the energy
spread is not the same at all energy points where data were collected,
the fake mass shift can appear.

To take into account the energy spread variations in the resonance
curve fit, we assumed that it linearly depends on the beam current and
the current density in the vicinity of their mean values
\begin{equation}
\begin{split}
 \sigma_E &\approx \left<\sigma_{E}\right>
    (1 +\alpha_i\!\cdot\! i + \alpha_j\!\cdot\! j),\\
  &
  i = \frac{I}{\left<I\right>}-1,\quad
  j = \frac{\mathcal{J}}{\left<\mathcal{J}\right>}-1.
\end{split}
\label{eqn:sigmaE}
\end{equation}
The product $I\!\cdot\!\Ilum^{\,n}$ with the specific luminosity
\mbox{$\Ilum = L/I_{+}I_{-} \propto 1/\sigma^{*}_x\sigma^{*}_y$} were
chosen as a measure of the current density effect as the beam sizes in
the ring were not permanently monitored.  The $\alpha_i$ and
$\alpha_j$ were considered as free parameters, the values $n=2$ and
$n=1$ were tried. If the synchrotron contribution in the radial size
dominates, and the vertical beam size is due to the coupling
($\sigma^{*}_y \propto \sigma^{*}_{x\beta}$), one has $n=2$ for the
multiple intra-beam scattering.  With $n=1$ the $j$ parameter just
characterizes the strength of the collision effects ($I\!\cdot\!\Ilum
\propto \xi_y$ for the flat beams).

\begin{figure}[tb]
\begin{center}
\includegraphics[clip,width=\columnwidth]{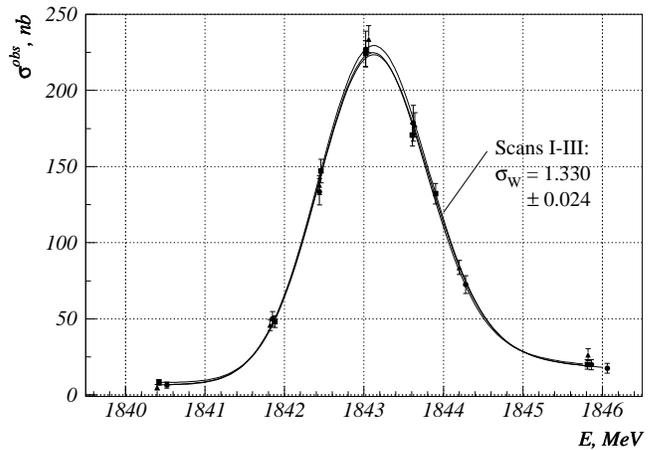}
\end{center}
\caption{\label{fig:psiprime} The results of three \PP
  scans (the mean energy spread value $\sigma_W$ is presented).  }
\end{figure}

\subsection{Fitting procedure}

Each data acquisition run was subdivided into subruns with a minor
variation of the beam currents. For each subrun the beam energy $E$
was assigned and the parameters $i_{\pm}$ and $j_{\pm}$ were
calculated for the $e^{+}$- and $e^{-}$-beams.

The observed number of the multihadron events $N$ has been fitted as
the function of $E$, $i$ and/or $j$ using the log-likelihood method.
The calculations of the expected number of the resonance events were
performed by the numerical convolution of (\ref{eqn:AZIMOV}) and
(\ref{eqn:Lunbiased}) with
\mbox{$\sigma_W=\sqrt{\sigma_{E+}^2+\sigma_{E-}^2}$} depending on
$i_{\pm}$ and/or $j_{\pm}$ according to (\ref{eqn:sigmaE}).  The free
parameters were: the constant \emph{continuum cross section}
$\sigma_c$, the event \emph{detection efficiency} $\epsilon$ for the
given value of the leptonic width $\Gamma_{ee}$, the \emph{resonance
  mass} $M$ and the \emph{energy spread parameters}
$\left<\sigma_W\right>$, $\alpha_i$ and/or $\alpha_j$. The PDG table
values \cite{PDG} were used for the resonance widths $\Gamma$ and
$\Gamma_{ee}$. The $\lambda$ parameter in (\ref{eqn:AZIMOV}) was fixed
as described above.

For the fitting procedure verification and the systematic error
checks, $N(E)$ fits were performed using (\ref{eqn:Dbeta}) with the
results presented in Fig.~\ref{fig:Jpsi} and Fig.~\ref{fig:psiprime}.

The peak height for the third \JP scan differs from those for the
first and the second ones due to the $\sigma_W$ dependence on the
parameter $j$. The energy spread in the fourth scan was decreased
intentionally (see Sec.~\ref{sec:Exp}).

In the \PP case, the energy spread variations are not seen
due to higher energy and narrower $i$-, $j$-ranges.

The chi-squared values of the fits are satisfactory
(\mbox{$P(\chi^2)>0.1$}) for all \JP and \PP scans even if the
dependence of the energy spread on the beam current within a scan is
ignored.
\begin{figure}[t]
\begin{center}
\includegraphics[clip,width=\columnwidth]{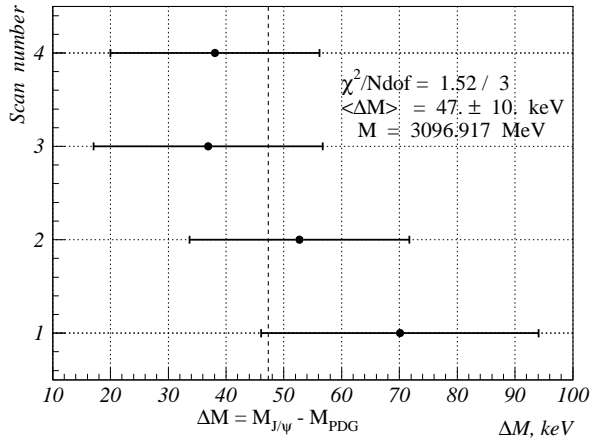}
\end{center}
\caption{\label{fig:MassJP}  \JP mass values
  for each scan relative to the world average value~\cite{PDG}, the
  resulting average values are shown by the dashed lines (statistical
  errors only).}
\end{figure}

The detection efficiencies obtained by the fit assuming the world
average values of the leptonic widths
\mbox{$\Gamma_{ee,\JP}=5.26\pm0.37$}~keV,
\mbox{$\Gamma_{ee,\PP}=2.19\pm0.15$}~keV
\cite{PDG} agree with
those obtained by Monte Carlo simulation within their errors.
The systematic error
of the Monte Carlo calculations and the error of the absolute luminosity
calibrations in this experiment are large (about 12\% in total), so we
do not present our leptonic width values. For the mass measurements
these errors are not important.

\section{Measured mass values and error discussion}\label{sec:ErrDisc}

\begin{figure}[t]
\begin{center}
\includegraphics[clip,width=\columnwidth]{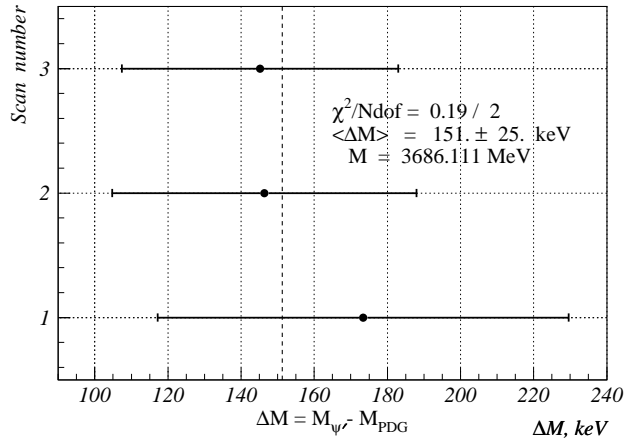}
\end{center}
\caption{\label{fig:MassPP}  \PP mass values
for each scan
relative to the world average value~\cite{PDG},
the resulting average values are shown by the dashed lines
(statistical errors only).}
\end{figure}

The mass values obtained in each scan \emph{assuming the constant value of $\sigma_W$ during a scan
}
and the resulting average mass values
for \JP and \PP
as well as their statistical errors are presented
in Figs.~\ref{fig:MassJP},\ref{fig:MassPP}.

To obtain the resulting averaged mass values, the scans were considered
as independent experiments. The individual mass values of the scans
were weighted using their statistical errors and ignoring the
systematic ones.
Correspondingly the systematic errors of the individual scans were weighted.
Such  procedure overestimates
the total error, but allows one to separate the statistical and
systematic errors of the resulting value.

The mass values for all scans are in good agreement even when the
difference in systematic errors is ignored (the $\chi^2$ is given 
in Figs.~\ref{fig:MassJP},\ref{fig:MassPP}).
The resulting statistical accuracy is \mbox{10~keV} for \JP and \mbox{25~keV}
for \PP.

The fit of \JP scans I$\div$III does not show the statistically
significant direct dependence on $I$
(\mbox{$\alpha_i\!=\!0.037\pm0.055$} ) and gives the dimensionless
slope \mbox{$\alpha_j\!=\!0.059\pm0.018$} similar to that of the
single-beam measurements.  The mass deviation from the value for
\mbox{$\alpha_i,\alpha_j =0$} does not exceed 1.5~keV.  The correction
is rather small, so the mass values for \mbox{$\alpha_i,\alpha_j =0$}
have been assigned to the scans I-III of \JP with the systematic error
of 1.5~keV related to the energy spread variations.
\begin{table*}[t]
\newcommand{\LT}{$<\,$}
\newcommand{\AP}{$\approx\,$}
\newcommand{\BL}{\phantom{$\approx\,$}}
\caption{\label{tab:Merr}
The systematic uncertainties in \JP and \PP masses~(keV)}
\begin{tabular}{p{0.6\textwidth}ll}\hline
\emph{Source} & \multicolumn{1}{c}{\mbox{$J/\psi$}} & 
                                  \multicolumn{1}{c}{\mbox{$\psi^{\prime}$}} \\
\hline
Energy spread variation  (Sec.~\ref{sec:ErrDisc},~\ref{sec:SigmaW}) 
                                                &\BL 3.0    &\BL2.0 \\
Energy assignment: statistical uncertainty 
                         (Sec.~\ref{sec:ErrDisc},~\ref{sec:Eass})
                                                &\BL 2.5    &\BL 3.5  \\
Energy assignment: prediction function choice
                         (Sec.~\ref{sec:ErrDisc},~\ref{sec:Eass})
                                                &\BL 2.7    &\BL 1.7  \\ 
Energy assignment: radial betatron oscillations (Sec.~\ref{sec:Eass})
                                                &\LT 1.5    &\LT 1.8  \\
Energy assignment: beam separation in the additional I.P. (Sec.~\ref{sec:Eass})
                                            &\BL 0.4$^{*}$   &\BL 0.4$^{*}$ \\
Beam misalignment in the interaction point (Sec.~\ref{sec:MeanW})
                                                &\BL 1.8     &\BL 5.1 \\
$e^{+}$-, $e^{-}$-energy difference (Sec.~\ref{sec:ErrDisc})
                                               &\LT 2.0 & \LT 2.0 \\
Non-gaussian collision energy distribution 
                               (Sec.~\ref{sec:ErrDisc},~\ref{sec:FitIntro})
                                               &\LT 1.5  &\LT 2.0 \\  
$\beta$-function chromaticity (Sec.~\ref{sec:MeanW})

                                   &\BL 2.0$^{*}$    &\BL 2.5$^{*}$ \\
Beam potential (Sec.~\ref{sec:MeanW})        &\BL 1.0$^{*}$  &\BL 1.0$^{*}$ \\
Single energy calibration (Sec.~\ref{sec:Cerr})  &\BL 0.6  &\BL  0.8 \\
Detection efficiency instability (Sec.~\ref{sec:EvSel})
                                              &\BL 2.3 &\BL  2.0 \\
Luminosity measurements  (Sec.~\ref{sec:LumMeas}) 
                                              &\BL 2.2 &\BL 3.0 \\
Interference in the hadronic channel  (Sec.~\ref{sec:Interf})
                                              &\BL 1.3 &\BL 0.8 \\
Residual machine background (Sec.~\ref{sec:EvSel})
                                              &\LT 1.0 &\LT1.0 \\
\hline
\emph{Sum in quadrature}                       &\AP 7.3  &\AP 8.9 \\ 
\hline
\multicolumn{3}{l}{$^{*}$~--- correction uncertainty}
\end{tabular}
\end{table*}
\begin{table*}[htb]
\caption{\label{tab:Mcorr}
The corrections applied to the \JP and \PP mass values~(keV).}
\begin{tabular}{p{0.44\textwidth}rrr}\hline
\emph{Correction for} 
        & \multicolumn{1}{c}{\mbox{$J/\psi$(I$\div$III)}}
                         & \multicolumn{1}{c}{\mbox{$J/\psi$(IV)}} &
                                 \multicolumn{1}{c}{\mbox{$\psi^{\prime}$}} \\
\hline
Vertical orbit disturbances (Sec.~\ref{sec:Cerr})
                            &$-0.8 \pm 0.6$ & $-0.8 \pm 0.6$ & $-0.6 \pm 0.4$ \\
Separation in the additional I.P. (Sec.~\ref{sec:Eass})
                            & $-3.4 \pm 0.4$ & $-3.4 \pm 0.4$ & $-4.0 \pm 0.4$ \\
$\beta$-function chromaticity (Sec.~\ref{sec:MeanW})
                            & $-4.0 \pm 2.0$ & $-1.5 \pm 0.7$ & $5.0 \pm 2.5$ \\
Beam potential (Sec.~\ref{sec:MeanW})
                            & $1.9 \pm 1.0$ & $2.1 \pm 1.0$ & $2.0 \pm 1.0$ \\
\hline
\emph{Total}                 & $-6.3 \pm 2.4$ & $-3.6 \pm 1.7$ & $1.4 \pm 2.8$ \\
\hline
\end{tabular}
\end{table*}

For the \JP scan IV with the damping decrements rearranged, one has
\mbox{$\alpha_j\!=\!0.45\pm0.15$} for $\alpha_i$ fixed at zero and
\mbox{$\alpha_i\!=\!-0.13\pm0.16$} for $\alpha_j$ fixed at zero with
the mass variations of $+10$~keV and $-8$~keV, respectively.  The mass
value for \mbox{$\alpha_i,\alpha_j =0$} has been assigned to the scan
IV of \JP with the systematic error of 10~keV related to the energy
spread variations.  The mass value obtained assuming the dependence on
\mbox{$J\propto I\!\cdot\!\Ilum^{\,1}$} does not contradict to this
error estimate.  The difference of the \JP mass values obtained in the
scan IV and in the scans I$\div$III is \mbox{($-11\pm22\pm12$)}~keV.

The systematic uncertainties are listed in Table~\ref{tab:Merr}.  The
sources of uncertainties not exceeding 0.3~keV 
(the uncertainties of the world average values of particle properties used,
the resonance mass uncertainty related to (\ref{eqn:Wmean}),
the radiative corrections uncertainties etc.)  are omitted.

The weighted contribution of the energy spread variation to the
systematic error of the resulting \JP
mass value is about 3~keV. 
During the \PP scans, the statistically
significant variations of the energy spread have not been observed, a
relatively big systematic error of $2$~keV is due to the bigger value of
$\sigma_W$.

The uncertainty of the energy assignment includes the statistical
component which was estimated using the multiple data fits with a
randomly generated energy deviation.  The systematic contribution
consists of two parts. The first one is related to the energy
prediction function (p.f.) choice and the energy assignment policy.
Three values of energy were tried for each run: the unbiased p.f.
value and two values of the shifted p.f. which exactly reproduce the
preceding and following calibration results. Besides, the best p.f.
parameter variation within their errors were allowed and a few
different functions were tried.  The second part is due to the
difference between the data acquisition runs and the energy
calibration runs (the radial betatron oscillations and the separation
in the additional I.P., Sec.~\ref{sec:Eass}).

The mass uncertainty caused by the beam misalignment in the
experimental I.P. and the electrostatic dispersion was evaluated
similarly to that of the statistical error of the energy assignment.

Estimates show that the difference of the energies of $e^{+}$- and
$e^{-}$-beams is small. The value shown in Table~\ref{tab:Merr} has
been obtained using a few energy calibrations with the $e^{+}$-beam
performed during the stability runs (Sec.~\ref{sec:Stab}).

The symmetric distortion of the distribution in the collision energy
$W$ can shift the mass due to the asymmetry of radiative corrections.
The corresponding uncertainty has been estimated by leaving the $k$
parameter of (\ref{eqn:Lunbiased}) free (the values shown in
Fig.~\ref{fig:MassJP},\ref{fig:MassPP} are for $k=0$).

All other sources of the systematic error were discussed above.  The
resulting systematic error in the mass is about 7~keV for \JP and
about 9~keV for \PP. The corrections applied to the fit results are
presented in Table~\ref{tab:Mcorr}.

The results obtained can be presented in the form
\begin{equation*}
  \begin{aligned}
   &M_{\JP} - M_{\JP}^{PDG} &= \,\ 47\pm 10\pm 7\ (\pm 40\ \text{\cite{PDG}}\,)\ \text{keV},\\
   &M_{\PP}\ -\ M_{\PP}^{PDG} &= 151\pm 25\pm 9\ (\pm 90\ \text{\cite{PDG}}\,)\ \text{keV},
  \end{aligned}
\end{equation*}
demonstrating the agreement with the world average values.

\section{Conclusion}

 The new high precision measurement of the \JP- and \PP-meson masses
has been performed at the collider \VEPP{4M} using the KEDR detector.
The following mass values have been obtained: 
\begin{equation*}
  \begin{aligned}
    &M_{\JP} &= 3096.917 &\pm 0.010 \pm 0.007\ \text{MeV} \\
    &M_{\PP} &= 3686.111 &\pm 0.025 \pm 0.009\ \text{MeV}
  \end{aligned}
\end{equation*}
The relative measurement accuracy reached \mbox{$4\cdot 10^{-6}$} for \JP,
\mbox{$7\cdot 10^{-6}$} for \PP and is approximately 3 times
better than that of the previous precise experiments. 

For the mass difference our result is
\[ M_{\psi^{\prime}} - M_{J/\psi} =  589.194 \pm 0.027 \pm 0.011\ \text{MeV.}\]

\section*{Acknowledgments}

We greatly appreciate the efforts of the staff of \VEPP{4M} to provide
good operation of the complex and the staff of experimental
laboratories for permanent support during preparation and performing
this experiment.  The authors express their special gratitude to
V.S.~Fadin and A.I.~Milstein for the theoretical support and
Yu.M.~Shatunov for stimulating discussions.

\bibliography{article}

\end{document}